\documentclass[sigconf,table, xcdraw, natbib]{acmart} 
\AtBeginDocument{%
  }

\setcopyright{acmlicensed}
\copyrightyear{2025}
\acmYear{2025}
\acmConference[CIKM '25] {Proceedings of the 34th ACM International Conference on Information and Knowledge Management}{ November 10--14, 2025}{Seoul, Republic of Korea.}
\acmBooktitle{Proceedings of the 34th ACM International Conference on Information and Knowledge Management (CIKM '25), November 10--14, 2025, Seoul, Republic of Korea}
\acmISBN{979-8-4007-2040-6/2025/11}
\acmDOI{10.1145/3746252.3761113}

\usepackage{graphicx} 
\usepackage{amsmath} 
\usepackage{booktabs}
\usepackage{caption}
\usepackage{makecell}
\usepackage{listings}
\usepackage[linesnumbered,ruled,vlined]{algorithm2e}
\usepackage{graphicx}
\usepackage{subcaption}
\usepackage{float} 
\usepackage{placeins}

\begin{document}

\title{{\textit{CLAP}: Coreference-Linked Augmentation for Passage Retrieval}}


\author{Huanwei Xu}
\affiliation{%
  \institution{School of Computer and Mathematical Sciences, The University of Adelaide}
  \city{Adelaide}
  \country{Australia}
}
\email{huanwei.xu@adelaide.edu.au}

\author{Lin Xu}
\affiliation{%
  \institution{School of Computer and Mathematical Sciences, The University of Adelaide}
  \city{Adelaide}
  \country{Australia}
}
\email{lin.xu@adelaide.edu.au}

\author{Liang Yuan}
\affiliation{%
  \institution{School of Computer and Mathematical Sciences, The University of Adelaide}
  \city{Adelaide}
  \country{Australia}
}
\email{liang.yuan@adelaide.edu.au}

\renewcommand{\shortauthors}{Huanwei Xu, Lin Xu, \& Liang Yuan}

\begin{abstract}

Large Language Model (LLM)-based passage expansion has shown promise for enhancing first-stage retrieval, but often underperforms with dense retrievers due to semantic drift and misalignment with their pretrained semantic space. Beyond this, only a portion of a passage is typically relevant to a query, while the rest introduces noise—an issue compounded by chunking techniques that break coreferential continuity. We propose \textbf{Coreference-Linked Augmentation for Passage Retrieval (CLAP)}, a lightweight LLM-based expansion framework that segments passages into coherent chunks, resolves coreference chains, and generates localized pseudo-queries aligned with dense retriever representations. A simple fusion of global topical signals and fine-grained subtopic signals achieves robust performance across domains. CLAP yields consistent gains even as retriever strength increases, enabling dense retrievers to match or surpass second-stage rerankers such as BM25 + MonoT5-3B, with up to 20.68\% absolute nDCG@10 improvement. These improvements are especially notable in out-of-domain settings, where conventional LLM-based expansion methods relying on domain knowledge often falter. CLAP instead adopts a logic-centric pipeline that enables robust, domain-agnostic generalization.

\end{abstract}




\maketitle

\section{Introduction}
First-stage passage retrieval is a fundamental component in open-domain question answering and information retrieval (IR) systems, where the goal is to efficiently identify relevant documents from large-scale corpora. Recent advances in dense retrieval, such as ANCE~\cite{ance} and SimLM~\cite{simlm}, have achieved strong performance by encoding queries and passages into a shared embedding space. However, these gains are often brittle in the face of out-of-domain generalization and semantically noisy passages, where retrieval effectiveness can significantly degrade.

One emerging solution to bridge the lexical-semantic gap is expansion-based matching, where queries or passages are augmented using large language models (LLMs). While intuitively appealing, recent studies~\cite{expansionfail} have shown that such expansions can degrade performance when combined with strong dense retrievers, particularly under distribution shifts. This degradation is largely attributed to a semantic mismatch between the generated expansions and the dense retriever’s pretrained embedding space. Moreover, as dense retrievers become stronger, they increasingly rely on precise semantic representations—making them more sensitive to noisy or spurious content introduced by expansions. As a result, instead of reinforcing true relevance, LLM-generated expansions may obscure it, leading to reduced effectiveness.

A key reason for this brittleness lies in the nature of current expansion strategies: they are typically \textit{knowledge-intensive}, relying heavily on the language model’s internalized domain knowledge to generate meaningful content. While this can be effective in-distribution, it poses a serious limitation when generalizing to unfamiliar domains. We argue for a shift toward a \textit{logic-intensive} paradigm—where expansions are grounded not in what the model knows about the world, but in what it can reason about from the surface structure of the passage. This logic-driven perspective enables expansion to be more controllable, interpretable, and robust across domains.

Beyond this representational mismatch, a deeper structural challenge persists: passages are seldom semantically uniform. Real-world passages frequently contain multiple subtopics, ambiguous coreference chains, and internal semantic drift. Naively expanding such passages in their entirety can amplify off-topic content and obscure the signals truly relevant to the query~\cite{expansionfail}.

To address these structural and representational limitations, we propose \textbf{Coreference-Linked Augmentation for Passage Retrieval (CLAP)}—a lightweight, LLM-based framework designed to improve first-stage dense retrieval. CLAP decomposes passages into semantically coherent units and augments them through a three-stage pipeline: (1) \textbf{semantic chunking} segments each passage into coherent subtopics, (2) \textbf{coreference resolution} clarifies ambiguous entity references within each chunk, and (3) a \textbf{prompt-driven generation module} produces localized pseudo-queries anchored to these resolved subtopics. While each module is lightweight by design, they form a principled framework that leverages LLMs' structural reasoning to improve retrieval effectiveness. To compute passage relevance, CLAP fuses two complementary signals: (i) the original query–passage similarity, and (ii) the maximum query–pseudo-query similarity across all localized pseudo-queries derived from the passage. A tunable fusion weight $\alpha$ balances global topical signals and fine-grained subtopic alignment.

The main contributions of this work are as follows:

\begin{itemize}
  \item We propose \textbf{CLAP}, a novel and modular LLM-based expansion framework that addresses both semantic and structural limitations of prior expansion methods for dense retrieval. Unlike knowledge-intensive expansion techniques, CLAP adopts a logic-driven approach grounded in the input’s internal structure.

  

  \item Unlike prior expansion methods such as HyDE, docT5query, and LameR—which generate global pseudo-queries or full-passage augmentations—CLAP introduce a three-stage expansion pipeline that integrates (1) \textbf{semantic chunking}, (2) \textbf{coreference resolution}, and (3) \textbf{multi-angle pseudo-query generation}, enabling fine-grained, subtopic-aware relevance modeling. This localized expansion better aligns with the retriever’s semantic space and avoids over-reliance on domain-specific knowledge.

  \item We design a lightweight fusion strategy that combines global and localized relevance signals through a tunable weight $\alpha$, yielding a robust and flexible first-stage ranking signal.

  \item We conduct extensive experiments on MS MARCO and four BEIR datasets, demonstrating that CLAP consistently improves both sparse and dense retrieval. On strong dense backbones such as SimLM and e5-large-v2, CLAP achieves gains up to +26.29 NDCG@10. On challenging out-of-domain datasets like ArguAna and FiQA, CLAP even \emph{outperforms} second-stage rerankers such as BM25$_{\text{top-100}}$ + MonoT5-3B~\cite{monot5}, underscoring the strength of logic-intensive, LLM-guided expansion.

\end{itemize}

The remainder of this paper is organized as follows. Section 2 reviews related work in passage expansion, semantic chunking, and LLM-based information retrieval. Section 3 introduces the proposed CLAP framework in detail, including its three-stage architecture and fusion-based relevance scoring strategy. While Section 4 presents our experimental setup, and Section 5 reports empirical results and ablation studies, highlighting the effectiveness of CLAP across diverse retrieval scenarios. Finally, Section 6 concludes with a discussion of limitations and future work.

\begin{figure}[t]
    \centering
    \includegraphics[width=\linewidth]{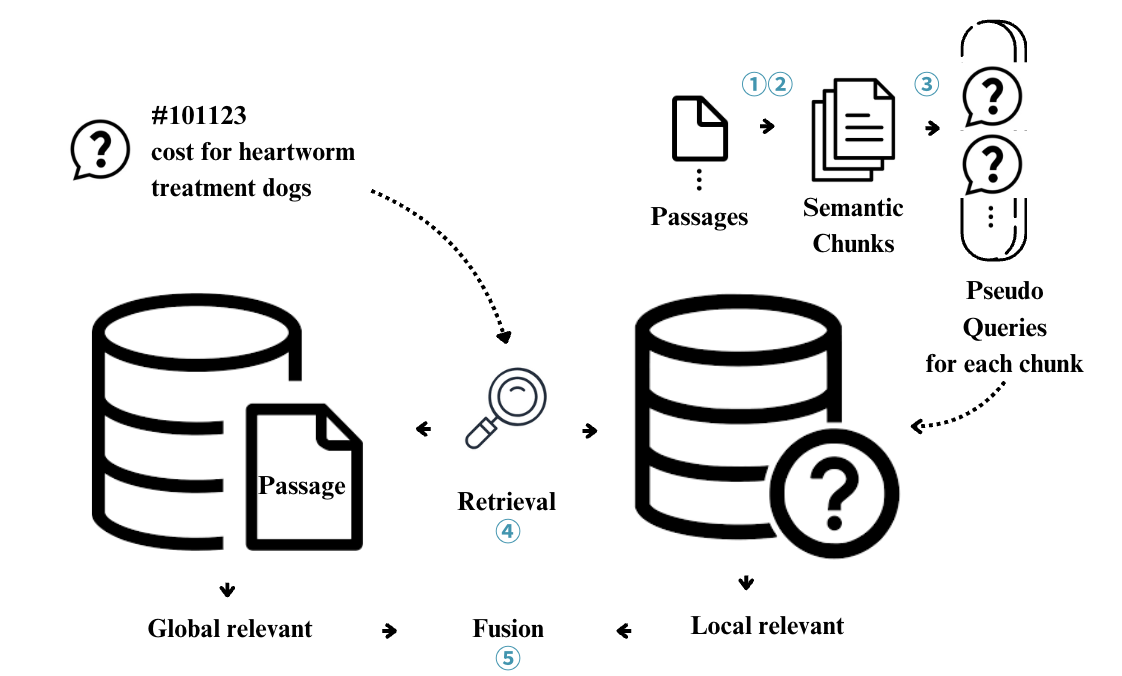}
    \caption{
    \textbf{CLAP’s modular five-stage pipeline.} 
(1) A passage is segmented into semantically coherent chunks. (2) Each chunk is coreference-resolved to ensure referential clarity. (3) Localized pseudo-queries are generated per chunk via prompt-based LLM inference. (4) Relevance is computed using both (a) query–passage similarity and (b) query–pseudo-query similarity. (5) Final passage ranking is produced by fusing global and local scores via interpolation.
    }
    \label{fig:pipeline}
\end{figure}

\section{Related Work}

\paragraph{Query and Passage Expansion.}
Expansion methods enhance retrieval effectiveness by enriching queries or passages with semantically related content. Traditional query expansion uses non-parametric signals like synonym sets or pseudo-relevance feedback~\cite{expansion1, expansion2, expansion3,cikm_p_prf}, while modern LLM-based approaches generate pseudo-passages (e.g., HyDE~\cite{hyde}) or augmentations such as Query2doc~\cite{query2doc} and LameR~\cite{lameR}. On the passage side, methods like docT5query~\cite{nogueira2019documentexpansionqueryprediction} produce pseudo-queries offline to support retrieval via denser index structures.

While these approaches have shown promise—particularly in sparse retrieval—LLM-based expansions often underperform in dense settings, especially under zero-shot or domain-shifted scenarios. This degradation is largely due to semantic misalignment between the generated expansions and the dense retriever’s pretrained embedding space~\cite{expansionfail}.  
That is, these global expansion strategies typically rely on the language model’s internalized domain knowledge to generate meaningful content, which may not generalize well to unfamiliar domains.

To address these limitations, CLAP introduces a contrasting, logic-driven approach to passage expansion. Rather than relying on domain knowledge implicitly encoded in the language model, CLAP grounds its expansion in the logical structure of the passage—segmenting it into coherent semantic units, resolving references, and generating diverse pseudo-queries that reflect fine-grained meaning. This structured, interpretable process enables more robust and domain-agnostic generalization than prior LLM-based expansion strategies.

\paragraph{Segmentation for Retrieval and CLAP's Positioning.}
Early passage retrieval leveraged fixed-size windows or paragraph splits to surface local evidence~\cite{kaszkiel1999efficient}. Discourse/topic segmentation sought linguistically-coherent boundaries (e.g., TextTiling~\cite{TextTiling}; minimum-cut~\cite{malioutov2006mincut}; linear segmentation~\cite{LinearTextSegmentation}), while diversification methods (MMR~\cite{mrr}, IA-Select~\cite{agrawal2009diversifying}, xQuAD~\cite{santos2010xquad}) improved \emph{result lists} by covering subtopics rather than restructuring documents. Modern dense retrieval typically keeps coarse, fixed passageing (ORQA~\cite{lee2019orqa}, DPR~\cite{karpukhin2020dpr}); late-interaction models capture fine-grained matches without document rewriting~\cite{khattab2020colbert}. In parallel, expansion approaches enrich \emph{global} representations (docT5query~\cite{nogueira2019documentexpansionqueryprediction}, SPLADE~\cite{formal2021splade}) or generate \emph{query-side} hypotheticals (HyDE~\cite{hyde}). By contrast, \textbf{CLAP} performs LLM-guided \emph{semantic chunking} and \emph{explicit coreference clarification} \emph{before} augmentation, yielding self-contained chunks and localized pseudo-queries, then fuses local and global signals—combining structural reorganization with targeted expansion. To our knowledge, this explicit pipeline of (i) semantic segmentation, (ii) cross-chunk disambiguation, and (iii) per-chunk expansion has not been jointly instantiated in prior IR work.

\paragraph{LLM-as-Agent Paradigm.}
Recent frameworks not only as text generators but also as agents capable of decomposing and solving complex tasks through modular reasoning pipelines. Frameworks like ReAct~\cite{react}, Toolformer~\cite{toolformer}, and CAMEL~\cite{camel} demonstrate that LLMs can coordinate multiple subtasks via prompt-driven planning, enabling tool use, multi-hop reasoning, and collaborative dialogue.

While prior agent-based methods focus on planning or dialogue tasks, CLAP applies this paradigm to first-stage retrieval. We design a lightweight multi-agent pipeline in which each subtask, including semantic chunking, coreference resolution, and pseudo-query generation, is delegated to an LLM using task-specific prompts. This structured orchestration leverages the decomposition capabilities of LLMs while maintaining simplicity and transparency in the expansion process.

\paragraph{Fusion-based Retrieval.}
Late interaction models like ColBERT~\cite{khattab2020colbert} and ColBERTv2~\cite{santhanam2022colbertv2} compute token-level similarity between queries and passages, achieving fine-grained matching at the cost of higher inference latency. SPLADE~\cite{formal2021splade} and SPLADE++~\cite{formal2022spladepp} adopt sparse lexical expansions that approximate dense matching while preserving compatibility with inverted indexing. Contriever~\cite{izacard2021contriever,cobertv2} and related reranking setups demonstrate that combining global dense retrieval with localized alignment can effectively bridge representational gaps.

CLAP instead achieves fine granularity by offloading complexity to the passage side through pseudo-query generation, allowing for efficient dual-view scoring with minimal inference overhead. Our findings align with hybrid models that show local-global fusion to be effective in diverse retrieval settings. Unlike many prior methods, CLAP requires \textit{no training }or\textit{ retriever finetuning}, making it easily applicable in zero-shot or resource-constrained scenarios.

\begin{figure}[t]
    \centering
    \includegraphics[width=\linewidth]{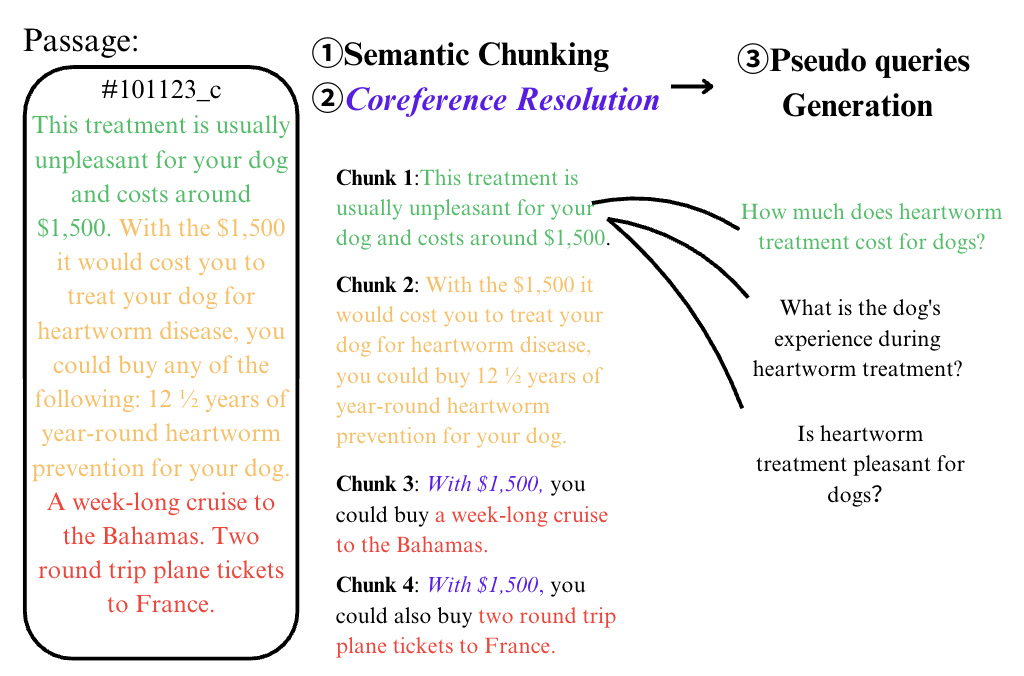}
    \caption{
    \textbf{Details of stages 1--3.} 
    (1) Semantic Chunking segments passages into coherent units. 
    (2) Coreference Resolution clarifies entity references. 
    (3) Pseudo-query Generation produces localized queries. 
    The query \#101123 is used as input. Green chunk: fully relevant; orange: partially relevant; red: irrelevant—illustrating semantic drift; purple: Coreference Resolution replaces pronouns with their explicit antecedents.
    }
    \label{fig:stages}
\end{figure}

\section{CLAP}
\label{sec:Method}

First-stage retrieval must balance two conflicting goals: retrieving passages that are broadly relevant at the topic level, and identifying fine-grained subtopic matches hidden within long, noisy documents. Directly matching entire passages to queries often amplifies semantic drift and coreferential ambiguity, while naive expansion risks overfitting to lexical noise.

To tackle these challenges, we propose \textbf{CLAP}, a lightweight pipeline that decomposes passages into semantically coherent units, resolves internal references, and generates localized pseudo-queries for retrieval. This design systematically introducing local (fine-grained) relevance signals by inducing a pseudo-query space that closely approximates the original query space. At the same time, we retain the dense retriever’s pretrained capacity as the global (coarse) relevance signals, and combining global and local relevance, enabling final scoring to capture both coarse and fine-grained signals in a unified view.

\subsection{Pipeline Overview}





CLAP consists of a lightweight, five-stage pipeline that enhances first-stage retrieval by decomposing passages and injecting fine-grained semantic signals. As shown in Figure~\ref{fig:pipeline}, the process is modular:

\begin{itemize}
    \item \textbf{Stage 1: Semantic Chunking.} The passage is segmented into coherent subtopic units.
    \item \textbf{Stage 2: Coreference Resolution.} Each chunk is made self-contained by replacing pronouns with their explicit antecedents.
    \item \textbf{Stage 3: Pseudo-query Generation.} An LLM generates multiple pseudo-queries per chunk, each targeting a distinct semantic perspective.
    \item \textbf{Stage 4: Dual Encoding.} All queries, passages, and pseudo-queries are embedded using a shared dense retriever.
    \item \textbf{Stage 5: Dual-view Retrieval.} Global (query–passage) and local (query–pseudo-query) scores are computed and combined via a tunable fusion weight $\alpha$.
\end{itemize}

Stages 1–3 convert noisy, multitopic passages into semantically enriched and disambiguated views, improving alignment with dense retrieval representations. Stages 4 to 5 then score passages from both global and localized perspectives, yielding robust retrieval that balances coarse and fine-grained relevance.

\vspace{0.5em}
\noindent\textbf{Notation.} We define the following key spaces and functions:
\begin{itemize}
    \item \(\mathcal{P}\): space of passages \(p_j \in \mathcal{P}\), encoded in \(\mathbb{R}^d\);
    \item \(\mathcal{Q}\): space of queries \(q_i \in \mathcal{Q} \subset \mathbb{R}^d\);
    \item \(\mathcal{C}\): space of semantic chunks, with \(\mathcal{C}(p)\) denoting the chunks from \(p\);
    \item \(\mathcal{T}\): space of chunk-level titles;
    \item \(\mathcal{PQ}\): space of pseudo-queries, where each \(\psi(\hat{c}_k)_l \in \mathcal{PQ}(\hat{c}_k)\) denotes the \(l\)-th pseudo-query generated from the coreference-resolved chunk \(\hat{c}_k \in \hat{\mathcal{C}}\), derived from original \(c_k \in \mathcal{C}\).

    \item \(\text{sim}(a,b)\): similarity function (e.g., cosine similarity);
    \item \(\operatorname{dist}(\mathcal{A},\mathcal{B})\): expected distance between embedding spaces.
\end{itemize}

\subsection{Semantic Chunking}
\label{sc}
Prior work has explored a variety of chunking strategies—ranging from fixed-size sliding windows~\cite{lee2019latent}, paragraph-based splits~\cite{karpukhin2020dense}, to topic-modeling guided segmentation~\cite{topic_model}—each offering trade-offs between semantic coherence and retrieval efficiency. However, these methods often fail to consistently isolate meaningful semantic units, especially when passages contain abrupt topical shifts.

Passages are often multitopical and exhibit internal semantic drift, for example, in the MS MARCO dataset (as shown on the left in Figure~\ref{fig:stages}), a ground truth passage for the query "\textit{cost for heartworm treatment dogs}" clearly provides relevant information, "\textit{This treatment is usually unpleasant for your dog and costs around \$1,500}", but subsequently introduces completely unrelated content, "\textit{A week-long cruise to the Bahamas. Two round trip plane tickets to France}", thus creating significant internal semantic interference.

To this end, we introduce \textit{semantic chunking} to transform a multitopic passage into semantically independent local relevant signals. To localize relevance signals, we first segment each passage \(p \in \mathcal{P}\) into coherent chunks using a semantic chunking agent \(\mathcal{G}_c\):
\[
\mathcal{G}_c(p_j) \rightarrow \{(c_k, \text{title}_k)\}_{k=1}^{K}
\]
where \(c_k \in \mathcal{C}\) is a subtopic and \(\text{title}_k \in \mathcal{T}\) summarizes its content.

Chunking ensures that subsequent expansions are grounded within well-defined semantic scopes.

\subsection{Coreference Resolution}
\label{cr}
Semantic chunking inevitably breaks cross-chunk context, where subsequent sentences may refer to previously mentioned entities via pronouns. Without resolving these references, the resulting chunks may become semantically incomplete or ambiguous—leading to degraded expansion quality and suboptimal retrieval performance. To ensure each chunk is self-contained, we apply coreference resolution using agent \(\mathcal{G}_r\):
\[
\mathcal{G}_r(c_k) \rightarrow \hat{c}_k
\]
where \(\hat{c}_k\) replaces pronouns and vague expressions in \(c_k\) with their explicit antecedents.

As shown in Figure~\ref{fig:stages}, coreference resolution strengthens the semantic independence of each chunk prior to pseudo-query generation.\footnote{We jointly perform semantic chunking and coreference resolution using a single prompt executed by an LLM. The complete prompt template is provided in Appendix~\ref{appendix:chunking_prompt}.}

\subsection{Pseudo-query Generation}
\label{pg}
For each resolved chunk \((\hat{c}_k, \text{title}_k)\), a pseudo-query generation agent \(\mathcal{G}_q\) generates multiple localized pseudo-queries:
\[
\mathcal{G}_q(\hat{c}_k, \text{title}_k) \rightarrow \{\psi(\hat{c}_k)_l\}_{l=1}^L
\]
yielding the pseudo-query set \(\mathcal{PQ}(\hat{c}_k)\).
\footnote{See Appendix~\ref{appendix:query_prompt} for the full prompt.}

These pseudo-queries aim to preserve the semantics of their source chunks while aligning closely with the query space \(\mathcal{Q}\), minimizing the expected representational distance:
\begin{equation}
\min_{\mathcal{G}_q} \; \mathbb{E} \left[ \operatorname{dist}(\mathcal{Q}, \mathcal{PQ}) \right]
\end{equation}

Together, they serve as a transformed representation of the original passage—where each pseudo-query captures a distinct semantic aspect (e.g., definition, causality, statistics. etc) within a specific subtopic (e.g., a semantic chunk)—thus acting as fine-grained, \textit{localized relevance} signals in the retrieval process (from \ref{sc} to \ref{pg}).

\begin{algorithm}[t]
\small
\SetAlgoNlRelativeSize{-2}
\caption{Local Relevance with Max-Pooling over Pseudo-Queries}
\label{alg:local-retrieval}
\KwIn{
Query embeddings \( \mathcal{Q} = \{q_1, \dots, q_{N_q}\} \subset \mathbb{R}^d \), \\
Pseudo-query embeddings \( \mathcal{PQ} = \{\psi_1, \dots, \psi_{N_l}\} \subset \mathbb{R}^d \), \\
Mapping \( m: \{1,\dots,N_l\} \rightarrow \{1,\dots,N_p\} \), \\
Top-$k$
}
\KwOut{Top-$k$ passages per query \( q_i \), stored in \( L[q_i] \)}

\textbf{Initialize:} local relevance dictionary \( L \gets \{\} \) \;

\ForEach{query \( q_i \in \mathcal{Q} \)}{
    \ForEach{pseudo-query \( \psi_l \in \mathcal{PQ} \)}{
        Let \( p_j = m(l) \) \tcp*{Parent passage of \(\psi_l\)}
        Let \( s = \text{sim}(q_i, \psi_l) \) \;
        \( L[q_i][p_j] \gets \max(L[q_i][p_j], s) \) \tcp*{Max over pseudo-queries}
    }
    Select top-$k$ passages from \( L[q_i] \) by descending score \;
}
\Return \( L \)

\vspace{0.5em}
\noindent \textbf{Note.}  \(m(l)\) denotes maps each pseudo-query to its parent passage.
\end{algorithm}

\begin{algorithm}[t]
\small
\SetAlgoNlRelativeSize{-2}
\caption{Relevance Score Fusion}
\label{alg:score-fusion}
\KwIn{
Global relevance scores \( G[q_i][p_j] \), \\
Local relevance scores \( L[q_i][p_j] \), \\
Weight \( \alpha \in [0,1] \), Top-$k$
}
\KwOut{Top-$k$ passages per query \( q_i \) ranked by fused score, \( F[q_i] \)}

\textbf{Initialize:} fused score dictionary \( F \gets \{\} \) \;

\ForEach{query \( q_i \in \mathcal{Q} \)}{
    \ForEach{passage \( p_j \in \mathcal{P} \)}{
        Let \( g \gets G[q_i][p_j] \), \( l \gets L[q_i][p_j] \) \;
        \( F[q_i][p_j] \gets \alpha \cdot g + (1 - \alpha) \cdot l \) \;
    }
    Select top-$k$ passages from \( F[q_i] \) by descending score \;
}
\Return \( F \)
\end{algorithm}

\subsection{Dual-View Retrieval}

Given a query \(q_i \in \mathcal{Q}\), we compute relevance at two levels:

\paragraph{Global Relevance.} Direct similarity between the query and passage:
\begin{equation}
\text{Global}(p_j \mid q_i) = \text{sim}(q_i, p_j)
\end{equation}

\paragraph{Local Relevance.} Maximum similarity between the query and any pseudo-query derived from the passage:

\begin{align}
\text{Local}(p_j \mid q_i) 
&= \max_{c_k \in \mathcal{C}(p_j)} \,
    \max_{\psi(c_k)_l \in \mathcal{PQ}(c_k)} \notag \\
&\quad \text{sim}\bigl(q_i, \psi(c_k)_l\bigr)
\end{align}

Local relevance captures fine-grained subtopic alignment often missed at the passage level. We outline the detailed procedure in Algorithm \ref{alg:local-retrieval}.

\subsection{Relevance Fusion}

We interpolate the two signals via a tunable coefficient \(\alpha\):
\begin{equation}
\begin{aligned}
\text{Final}(p_j \mid q_i) &= \alpha \cdot \text{Global}(p_j \mid q_i) \\
&\quad + (1 - \alpha) \cdot \text{Local}(p_j \mid q_i)
\end{aligned}
\end{equation}
where \(\alpha \in [0,1]\) balances coarse and fine-grained relevance. We tune \(\alpha\) per dataset. Algorithm \ref{alg:score-fusion} shows the implementation.

\noindent \textbf{Summary.} Unlike prior expansion methods that directly modify query or passage content—often at the risk of semantic drift—CLAP enhances retrieval by structurally transforming the passage into localized signals while preserving the original dense retriever’s semantic space. Through dual-view scoring and lightweight fusion, it captures both broad topical alignment and fine-grained relevance, offering a generalizable and training-free augmentation strategy across domains.



\begin{table}[t]
\centering
\scriptsize
\setlength{\tabcolsep}{8pt}
\renewcommand{\arraystretch}{1}
\begin{tabular}{l|c|ccc}
\toprule
\textbf{Method} & \textbf{Size/Dim} & \multicolumn{3}{c}{\textbf{MS MARCO dev}} \\
\cmidrule(lr){3-5}
& & MRR@10 & R@1k & nDCG@10 \\
\midrule
\multicolumn{5}{l}{\textbf{Sparse Retrieval}} \\
BM25 & - & 30.88 & 92.56 & 38.01 \\
\quad + query2doc & - & \cellcolor{blue!15}39.42 & \cellcolor{blue!15}93.75 & \cellcolor{blue!15}46.93 \\
\quad + \textbf{CLAP}\textsubscript{ ($\alpha = 0.2$)} & - & \cellcolor{blue!15}\textbf{41.55}\textsuperscript{+10.67} & \cellcolor{blue!15}\textbf{95.61}\textsuperscript{+3.05} & \cellcolor{blue!15}\textbf{47.41}\textsuperscript{+9.40} \\
\quad \quad + both* \textsubscript{ ($\alpha = 0.1$)} & - & \cellcolor{blue!15}44.23\textsuperscript{+13.35} & \cellcolor{blue!15}97.19\textsuperscript{+4.63} & \cellcolor{blue!15}50.57\textsuperscript{+12.56} \\
\midrule
\multicolumn{5}{l}{\textbf{Dense Retrieval w/o Distillation}} \\
ANCE & 110M / 768 & 48.84 & 98.85 & 56.52 \\
\quad + query2doc & - & \cellcolor{red!15}48.46 & \cellcolor{red!15}98.49 & \cellcolor{red!15}55.76 \\
\quad + \textbf{CLAP} \textsubscript{ ($\alpha = 0.3$)}& - & \cellcolor{blue!15}\textbf{53.56}\textsuperscript{+4.72} & \cellcolor{blue!15}\textbf{99.39}\textsuperscript{+0.54} & \cellcolor{blue!15}\textbf{59.16}\textsuperscript{+2.64} \\
\quad \quad + both* \textsubscript{ ($\alpha = 0.5$)} & - & \cellcolor{blue!15}54.56\textsuperscript{+5.72} & \cellcolor{blue!15}98.97\textsuperscript{+0.12} & \cellcolor{blue!15}60.05\textsuperscript{+3.53} \\
\midrule
\multicolumn{5}{l}{\textbf{Dense Retrieval w/ Distillation}} \\
SimLM & 336M / 768 & 51.74 & \textbf{98.92} & 59.00 \\
\quad + query2doc & - & \cellcolor{red!15}48.86 & \cellcolor{red!15}98.41 & \cellcolor{red!15}55.93 \\
\quad + \textbf{CLAP} \textsubscript{ ($\alpha = 0.4$)}& - & \cellcolor{blue!15}\textbf{54.21}\textsuperscript{+2.47} & \textbf{98.92}\textsuperscript{+0.00} & \cellcolor{blue!15}\textbf{59.77}\textsuperscript{+0.77} \\
\quad \quad + both* \textsubscript{ ($\alpha = 0.4$)}& - & \cellcolor{blue!15}55.37\textsuperscript{+3.63} & \cellcolor{red!15}98.00\textsuperscript{-0.92} & \cellcolor{blue!15}60.64\textsuperscript{+1.64} \\
all-MiniLM-L6-v2 & 22M / 384 & 49.70 & 99.46 & 57.73 \\
\quad + query2doc & - & \cellcolor{red!15}49.56 & \cellcolor{red!15}99.28 & \cellcolor{red!15}57.10 \\
\quad + \textbf{CLAP}\textsubscript{ ($\alpha = 0.3$)} & - & \cellcolor{blue!15}\textbf{52.68}\textsuperscript{+2.98} & \cellcolor{blue!15}\textbf{99.63}\textsuperscript{+0.17} & \cellcolor{blue!15}\textbf{58.72}\textsuperscript{+0.99} \\
\quad \quad + both* \textsubscript{ ($\alpha = 0.3$)} & - & \cellcolor{blue!15}53.11\textsuperscript{+3.41} & \cellcolor{red!15}91.35\textsuperscript{-8.11} & \cellcolor{blue!15}59.07\textsuperscript{+1.34} \\
e5-large-v2 & 335M / 1024 & 54.92 & \textbf{99.68} & \textbf{62.62} \\
\quad + query2doc & - & \cellcolor{red!15}50.71 & \cellcolor{red!15}99.12 & \cellcolor{red!15}58.00 \\
\quad + \textbf{CLAP} \textsubscript{ ($\alpha = 0.4$)}& - & \cellcolor{blue!15}\textbf{56.42}\textsuperscript{+1.50} & \cellcolor{red!15}99.52\textsuperscript{-0.16} & \cellcolor{red!15}62.02\textsuperscript{-0.60} \\
\quad \quad + both* \textsubscript{ ($\alpha = 0.3$)}& - & \cellcolor{blue!15}55.00\textsuperscript{+1.58} & \cellcolor{red!15}99.44\textsuperscript{-0.24} & \cellcolor{red!15}60.80\textsuperscript{-1.82} \\
\bottomrule
\end{tabular}
\caption{Performance on the MS MARCO dev set. 
\textbf{Size/Dim} denotes the number of model parameters and embedding dimensionality. 
\textbf{Blue shading} indicates positive performance gains, while \textbf{red shading} indicates degradation. 
\textbf{+both} refers to the simultaneous use of query and passage expansion. 
The results for \textbf{query2doc} are reproduced by us.}
\label{tab:msmarco-ours}
\end{table}

\section{Experiments}

\begin{table*}[t]
\centering
\small
\setlength{\tabcolsep}{10pt}
\renewcommand{\arraystretch}{1}
\begin{tabular}{l|c|cccc|c}
\toprule
\textbf{Method} & \textbf{Size/Dim} & \textbf{SciFact} & \textbf{ArguAna} & \textbf{FiQA} & \textbf{NFCorpus} & \textbf{Avg Gain} \\
\midrule
\multicolumn{7}{l}{\textbf{Sparse Retrieval}} \\
BM25 & - & 66.28 & 31.43 & 23.60 & 30.67 & - \\
\quad + docT5query\cite{nogueira2019documentexpansionqueryprediction} & - & \cellcolor{blue!15}67.50 & \cellcolor{blue!15}34.90 & \cellcolor{blue!15}29.10 & \cellcolor{blue!15}32.80 & - \\
\quad + query2doc\cite{query2doc} & - & \cellcolor{blue!15}68.60 & - & - & \cellcolor{blue!15}\textbf{37.00} & - \\
\quad + LameR\cite{lameR} & - & \cellcolor{blue!15}\textbf{73.50} & \cellcolor{blue!15}40.20 & \cellcolor{blue!15}25.80 & - & - \\
\quad + \textbf{CLAP} \textsubscript{ ($\alpha$ = \{0.6, 0.3, 0.2, 0.8\}) } & - & \cellcolor{blue!15}71.05\textsuperscript{+4.77} & \cellcolor{blue!15}\textbf{40.56}\textsuperscript{+9.13} & \cellcolor{blue!15}\textbf{52.38}\textsuperscript{+28.78} & \cellcolor{blue!15}31.40\textsuperscript{+0.73} & \cellcolor{blue!15}+10.85 \\
\midrule
\multicolumn{7}{l}{\textbf{Dense Retrieval w/o Distillation}} \\

ANCE & 110M / 768 & 51.14 & 41.89 & 29.49 & 23.57 & - \\
\quad + \textbf{CLAP} \textsubscript{ ($\alpha$ = \{0.2, 0.4, 0.0, 0.4\}) } & - & \cellcolor{blue!15}\textbf{64.88}\textsuperscript{+13.74} & \cellcolor{blue!15}\textbf{53.86}\textsuperscript{+11.97} & \cellcolor{blue!15}\textbf{55.78}\textsuperscript{+26.29} & \cellcolor{blue!15}\textbf{28.84}\textsuperscript{+5.27} & \cellcolor{blue!15}+14.32 \\

HyDE\cite{hyde} & - & \cellcolor{blue!15}69.1 & \cellcolor{blue!15}46.6 & \cellcolor{red!15}27.3 & -  & - \\
\midrule
\multicolumn{7}{l}{\textbf{Dense Retrieval w/ Distillation}} \\
SimLM & 336M / 768 & 61.27 & 36.05 & 28.77 & 32.37 & - \\
\quad + query2doc\cite{query2doc} & - & \cellcolor{red!15}59.50 & - & - & \cellcolor{red!15}32.10 & - \\
\quad + \textbf{CLAP} \textsubscript{ ($\alpha$ = \{0.0, 0.3, 0.0, 0.6\})  }& - & \cellcolor{blue!15}\textbf{69.75}\textsuperscript{+8.48} & \cellcolor{blue!15}\textbf{53.88}\textsuperscript{+17.83} & \cellcolor{blue!15}\textbf{54.86}\textsuperscript{+26.09} & \cellcolor{blue!15}\textbf{34.92}\textsuperscript{+2.55} & \cellcolor{blue!15}+13.74 \\
all-MiniLM-L6-v2  & 22M / 384 & 64.00 & 50.15 & 36.87 & 31.15 & - \\
\quad + \textbf{CLAP} \textsubscript{ ($\alpha$ = \{0.0, 0.3, 0.2, 0.3\}) }& - & \cellcolor{blue!15}\textbf{71.66}\textsuperscript{+7.66} & \cellcolor{blue!15}\textbf{58.11}\textsuperscript{+7.96} & \cellcolor{blue!15}\textbf{59.96}\textsuperscript{+23.09} & \cellcolor{blue!15}\textbf{35.55}\textsuperscript{+4.40} & \cellcolor{blue!15}+10.78 \\
e5-large-v2 & 335M / 1024 & 70.76 & 51.77 & 37.11 & 35.60 & - \\
\quad + \textbf{CLAP} \textsubscript{ ($\alpha$ = \{0.0, 0.4, 0.0, 0.6\}) } & - & \cellcolor{blue!15}\textbf{74.72}\textsuperscript{+3.96} & \cellcolor{blue!15}\textbf{63.08}\textsuperscript{+11.31} & \cellcolor{blue!15}\textbf{59.15}\textsuperscript{+22.04} & \cellcolor{blue!15}\textbf{38.43}\textsuperscript{+2.83} & \cellcolor{blue!15}+10.04 \\
\midrule
\multicolumn{7}{l}{\textbf{Re-ranking}\textsubscript{w/o expansion}} \\

BM25\textsubscript{$top100$} + MonoT5-3B\cite{expansionfail} & 3B/ - & 82.10 & 42.40 & 45.90 & 39.20 & - \\
\bottomrule

\end{tabular}
\caption{\textbf{NDCG@10 results on selected BEIR datasets.} 
\textbf{Avg Gain} reports the average improvement across datasets for each model configuration. ($\alpha$ = \{w,x,y,z\}) represent $\alpha$ setting of each dataset from left to right.
Other annotations follow the same conventions as Table~\ref{tab:msmarco-ours}.}
\label{tab:beir-results-updated}
\end{table*}

\subsection{Evaluation Setup}

\paragraph{Datasets and Metrics.}
We evaluate CLAP on both in-domain and zero-shot out-of-domain settings. In-domain evaluation is conducted on the MS MARCO dev set\cite{msmacro}, a large-scale passage retrieval benchmark. For out-of-domain generalization, we use four diverse datasets from BEIR\cite{beir2021}: (1) SciFact (scientific fact checking), (2) ArguAna (argument retrieval), (3) FiQA-2018 (financial QA), and (4) NFCorpus (biomedical). We report MRR@10, Recall@1000, and nDCG@10 for MS MARCO, and nDCG@10 for BEIR.

\paragraph{Baselines and Comparisons.}
We compare CLAP with strong baselines: \\ \textbf{BM25}~\cite{bm25} (sparse), \textbf{ANCE}~\cite{ance} (dense w/o distillation), and three distillation-based dense retrievers: \textbf{SimLM}~\cite{simlm}, \textbf{MiniLM-L6-v2}~\cite{minilm}, and \textbf{e5-large-v2}~\cite{e5-large}. For expansion-based baselines, we reproduce \textbf{Query2doc}~\cite{query2doc} on MS MARCO, and compare with \textbf{docT5query}~\cite{nogueira2019documentexpansionqueryprediction} , \textbf{HyDE}~\cite{hyde} and \textbf{LameR}~\cite{lameR} on BEIR. We also include \textbf{BM25 + MonoT5-3B}~\cite{monot5} as a strong reranker upper bound.

\paragraph{Implementation Details.}
We apply CLAP offline to MS MARCO dev passages (484k) due to compute limits. We use \textbf{Mistral-Large-24.11}~\cite{mistral2024} with \textit{temperature 0} for chunking and pseudo-query generation. Passages over 5000 words are skipped from chunking and directly used. BM25 is implemented with Elasticsearch~\cite{elasticsearch} using BEIR hyperparameters. Evaluation is performed with \texttt{pytrec\_eval}~\cite{Pytrec_eval}.

We provide all code, processed data, and experimental scripts in repository: \url{https://github.com/iamulaodou/CLAP}.

\begin{figure}[t]
  \centering
  \begin{subfigure}{0.35\textwidth}
    \centering
    \includegraphics[width=\linewidth]{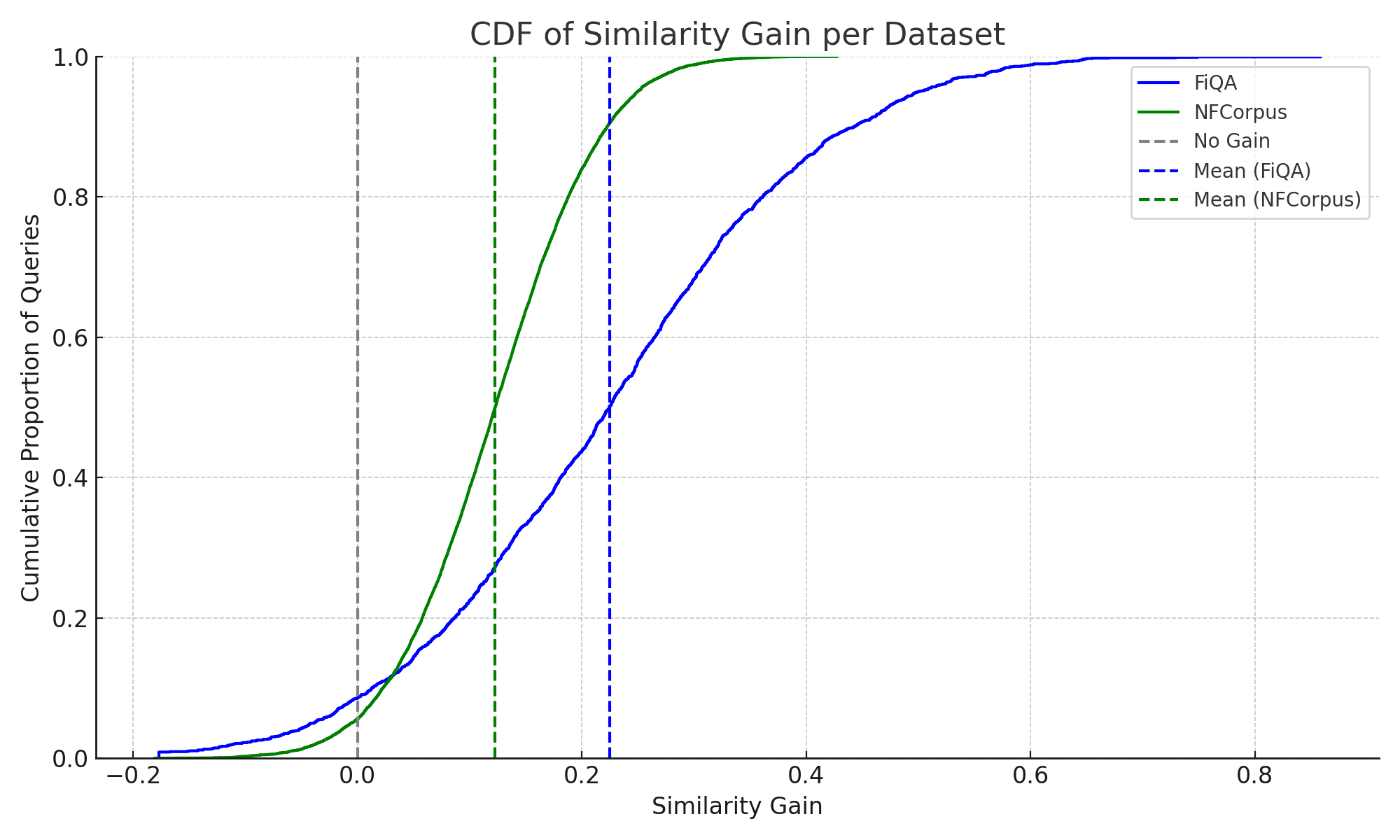}
    \caption{Cumulative distribution of similarity gain across queries on SciFact and FiQA datasets. Dashed lines indicate the mean improvement per dataset.}
    \label{fig:cdf-gain}
  \end{subfigure}
  \hfill
  \begin{subfigure}{0.35\textwidth}
    \centering
    \includegraphics[width=\linewidth]{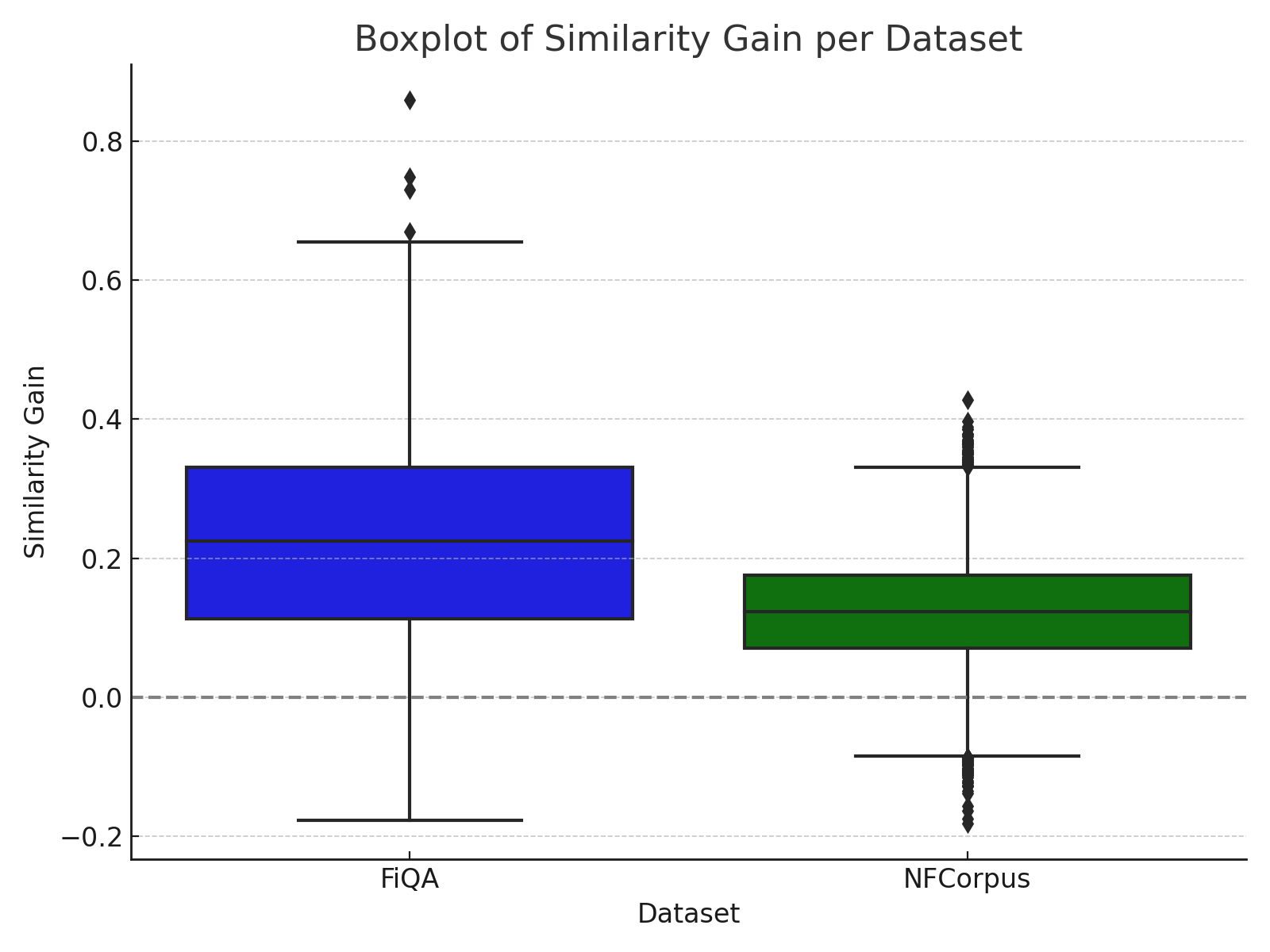}
    \caption{Box plot of similarity gain per dataset. Dashed line at 0 indicates no gain.}
    \label{fig:box-gain}
  \end{subfigure}
  \caption{Similarity gain analysis across datasets.}
  \label{fig:gain-combined}
\end{figure}

\subsection{Main Results}

\paragraph{In-domain Performance.}
Table~\ref{tab:msmarco-ours} shows that CLAP consistently improves retrieval performance across all retrievers on MS MARCO. Gains are especially large for sparse retrievers: BM25 sees +10.67 MRR and +9.40 nDCG. For dense retrievers, gains are smaller but consistent. Notably, combining CLAP with query2doc—i.e., using both pseudo-query-based local relevance and query-expanded global relevance—yields further improvements. This highlights that stronger global signals (e.g., from lexical expansions by query2doc) provide a better foundation for local signal fusion, amplifying the effectiveness of localized retrieval cues generated by CLAP.

\paragraph{Out-of-domain Generalization.}
Table~\ref{tab:beir-results-updated} presents results on BEIR. CLAP yields strong improvements across both sparse and dense retrievers. For instance, it brings +28.78 nDCG@10 gain on FiQA with BM25 and an average gain of +14.32 on ANCE. Compared to expansion methods like query2doc—and as previously observed for docT5query~\cite{expansionfail}—which often degrade performance when paired with strong dense retrievers, CLAP consistently delivers improvements. When compared to HyDE~\cite{hyde}, CLAP attains higher nDCG@10 scores on ArguAna (53.86 vs.\ 46.6) and FiQA (55.78 vs.\ 27.3), while being 4.22 points lower on SciFact (64.88 vs.\ 69.1). Notably, CLAP achieves average gains of +13.74 on SimLM, +10.78 on MiniLM, and +10.04 on e5-large, indicating that its effectiveness remains stable even when applied to more powerful dense retrievers. This robustness suggests that CLAP is retriever-agnostic and capable of stably raising the performance upper bound across architectures and model scales. Moreover, CLAP combined with e5-large matches or exceeds second-stage rerankers (e.g., BM25 + MonoT5-3B) on ArguAna and FiQA, showcasing its strong potential as a first-stage retriever enhancement.

\begin{table}[t]
\centering
\small
\setlength{\tabcolsep}{6pt}
\renewcommand{\arraystretch}{1.2}
\begin{tabular}{l|c}
\toprule
\textbf{Setting\textsubscript{(all-MiniLM-L6-v2)}} & \textbf{NDCG@10 \textsubscript{(ArguAna)}} \\
\midrule
Baseline (Query–Passage) & 50.15 \\
Query–Chunk w/o Coref & 35.77 \\
Query–PseudoQuery w/o Coref & 35.72 \\

\midrule
Query–Chunk w/ Coref & 36.33  \\
Query–PseudoQuery w/ Coref & \textbf{54.61} \\
\midrule
Fusion ($\alpha = 0.3$) & \textbf{58.11} \\

\bottomrule
\end{tabular}
\caption{Ablation on pseudo-query source and coreference resolution. Fusion denotes weighted interpolation between original passage score and pseudo-query score.}
\label{tab:ablation-arguana}
\end{table}

\begin{table*}[t]
\centering
\small
\setlength{\tabcolsep}{2pt}
\renewcommand{\arraystretch}{1.2}
\begin{tabular}{lcccccccccc}
\toprule
\textbf{Dataset} 
& \makecell{\textbf{Avg.}\\\textbf{Query Len}} 
& \makecell{\textbf{Avg.}\\\textbf{Passage Len}} 
& \makecell{\textbf{Len. Ratio}\\\textbf{(P/Q)}} 
& \makecell{\textbf{Avg.}\\\textbf{P/Q}} 
& \makecell{\textbf{Avg.}\\\textbf{$\alpha$}} 
& \textbf{BM25} 
& \textbf{ANCE} 
& \textbf{SimLM} 
& \makecell{\textbf{MiniLM}\\\textbf{-L6}} 
& \makecell{\textbf{e5}\\\textbf{-Large}} \\
\midrule
NFCorpus   & 3.30   & 232.26 & 70.38 & 38.2 & 0.54 & 0.8 & 0.4 & 0.6 & 0.3 & 0.6 \\
SciFact    & 12.37  & 213.63 & 17.27 & 1.1  & 0.22 & 0.6 & 0.2 & 0.2 & 0.1 & 0.0 \\
ArguAna    & 192.98 & 166.80 & 0.86  & 1.0  & 0.34 & 0.3 & 0.4 & 0.3 & 0.3 & 0.4 \\
FiQA       & 10.77  & 132.32 & 12.29 & 1.1  & 0.08 & 0.2 & 0.0 & 0.1 & 0.1 & 0.0 \\
\bottomrule
\end{tabular}

\caption{Average query/passage length, their ratio and optimal fusion weights ($\alpha$) across datasets and retrievers. A higher $\alpha$ indicates greater reliance on global (query-to-passage) relevance, while lower values emphasize local (query-pseudo query) relevance(see figure \ref{fig:viz-alpha}, appendix \ref{appendix:appendix_alpha} for full setting). Avg. P/Q
indicates the average relevant passages per query.}
\label{tab:alpha_analysis}
\end{table*}

\subsection{Ablation and Analysis}
\paragraph{Fine-Grained Relevance Similarity Gains.}
\label{analysis}
First, We define the similarity gain as the difference between the best pseudo-query match and the original \textit{ground-truth} passage similarity for particular query:
\begin{equation}
\text{Similarity Gain} = \max_{\psi \in \mathcal{PQ}(p)} \text{sim}(q, \psi) - \text{sim}(q, p)
\end{equation}

This value quantifies the net benefit of localized pseudo-query matching for each ground-truth passage. Figures~\ref{fig:cdf-gain} and~\ref{fig:box-gain} visualize this gain across FiQA and NFCorpus on all-MiniLM-L6-v2.

Both datasets show that over 90\% of queries achieve positive gains. In \textbf{FiQA}, the median similarity gain is $+0.19$, and the mean is $+0.22$, indicating consistent and substantial improvements. Meanwhile, \textbf{NFCorpus} yields a median of $+0.12$ and a mean of $+0.12$—still positive but more conservative. The range of gains is also wider in FiQA (from $-0.18$ to $+0.94$), compared to NFCorpus ($-0.27$ to $+0.55$). The lower standard deviation in NFCorpus ($0.078$ vs.\ $0.167$) reflects this tighter distribution. For detailed descriptive statistics, please see the Appendix ~\ref{appendix:similarity_stats}. 

Despite both showing similarity-level improvements, their impact on retrieval metrics diverges sharply. FiQA shows the largest retrieval boost among all BEIR datasets, while NFCorpus sees only modest gains. This contrast stems from corpus characteristics: FiQA's question-answer format aligns well with chunk-level pseudo-queries, amplifying their effect. In contrast, NFCorpus presents a challenging setting with short, vague queries (3.3) and long biomedical passages (232.26), coupled with a high ground-truth multiplicity—where a large number of passages are labeled relevant per query  (38.2). In such cases, similarity gains for a single ground-truth passage have limited impact on overall ranking metrics, diluting the effect of localized improvements.

\paragraph{Coreference Resolution.} Table~\ref{tab:ablation-arguana} examines the role of coreference resolution on ArguAna. Without coref, pseudo-queries hurt performance (35.72 vs. 50.15 baseline). With coref, pseudo-query matching improves to 54.61. Fusion of global and local scores (\(\alpha=0.3\)) further boosts to 58.11, confirming that resolved, self-contained chunks are critical to CLAP's effectiveness.


\paragraph{Fusion weight \boldmath$\alpha$ and when CLAP helps.}
Table~\ref{tab:alpha_analysis} reveals \emph{three corpus regimes} rather than a single monotone trend.

(1) \textbf{NFCorpus (extremely short queries; high length ratio).}
Queries are very short while passages are long (length ratio \( \mathrm{P}/\mathrm{Q}=70.38 \)).
Under this mismatch, the local chunk→pseudo signal does not cleanly dominate the global
query–passage signal; both are diluted by abundant, lengthy context. The optimal \( \alpha \)
is therefore \emph{large} (avg.\ 0.54; up to 0.8 for some retrievers), placing more weight on
global matching. Notably, in this regime \emph{all expansion methods} in our experiment yield only marginal gains.
(2) \textbf{SciFact \& FiQA (localized evidence)}
With medium queries and moderate Len. Ratio (P/Q), the informative evidence is concentrated in a small span. Here the local signal  \emph{clearly separates} true passages and mitigates passage-level noise. Consequently the optimal $\alpha$ is \emph{small} (avg.\ 0.22 and 0.08, respectively), and we observe a consistent tendency for \emph{stronger encoders to prefer even smaller $\alpha$}, as they better exploit fine-grained local semantics.
(3) \textbf{ArguAna (extreme long query).}
After CLAP’s pipeline, retrieval effectively becomes answer(the long query) finds question(the processed pseudo-query). Local matching still dominates, but the \emph{global component matters more than in Regime (2)}; the best $\alpha$ is \emph{moderate} (avg.\ 0.34).

\textit{Practical guidance.}
Inspect a tiny dev slice (50–100 queries). If queries are extremely short(Regime 1), start with $\alpha\in[0.5,0.8]$; if evidence is localized(Regime 2), start with $\alpha\in[0.0,0.2]$ (lower for stronger encoders); if queries are longer than passages (Regime 3), use a moderate range $\alpha\in[0.25,0.40]$. A 5–7 point grid around the initial setting suffices. Figure~\ref{fig:viz-alpha} (Appendix~\ref{appendix:appendix_alpha}) further visualizes these trends across \(\alpha \in [0.0,1.0]\).



\begin{table}[t]
\centering
\begin{minipage}{0.48\linewidth}
\centering
\small
\setlength{\tabcolsep}{4pt}
\renewcommand{\arraystretch}{1.2}
\begin{tabular}{l|c}
\toprule
\textbf{Temperature} & \makecell{\textbf{NDCG@10}\\\textbf{(SciFact, $\alpha=0$)}} \\
\midrule
0.0 & 71.667 \\
0.5 & 66.666 \\
\bottomrule
\end{tabular}
\caption{Effect of LLM temperature settings during pseudo-query generation. Lower temperature (more deterministic outputs) leads to better retrieval performance.}
\label{tab:temperature-study}
\end{minipage}
\hfill
\begin{minipage}{0.48\linewidth}
\centering
\small
\setlength{\tabcolsep}{3pt}
\renewcommand{\arraystretch}{1.1}
\begin{tabular}{lccc}
\toprule
\textbf{Dataset} & $p/q$ & $c/p$ & $pq/c$ \\
\midrule
SciFact   & 17.28 & 6.94 & 4.50 \\
ArguAna   &  6.17 & 5.63 & 3.32 \\
FiQA      & 88.95 & 5.00 & 4.58 \\
NFCorpus  & 11.25 & 6.77 & 4.70 \\
\bottomrule
\end{tabular}
\caption{Structural statistics of the CLAP preprocessing pipeline.
$p/q$ = avg.\ passage-to-query length ratio,  
$c/p$ = avg.\ chunks per passage,  
$pq/c$ = avg.\ pseudo-queries per chunk.}
\label{tab:structure-stats}
\end{minipage}
\end{table}

\paragraph{Effect of Temperature.} Table~\ref{tab:temperature-study} evaluates decoding temperature. Lower temperature (0.0) outperforms higher (0.5), suggesting that diversity harms pseudo-query fidelity. Since each chunk is already semantically distinct, diverse outputs often introduce irrelevant drift.

\paragraph{Index Footprint and LLM Preprocessing Cost.}
As Table~\ref{tab:structure_stats} shows, CLAP exhibits a consistent expansion profile across BEIR datasets: each passage is segmented into $c/p \approx 5\text{--}7$ chunks, and each chunk yields $pq/c \approx 4\text{--}5$ pseudo-queries. As a result, the retrieval index grows by roughly $20\text{--}32\times$ compared to the original corpus. While this expansion still preserves linear-time retrieval complexity, it incurs a notable increase in vector storage and computation. Nonetheless, the consistency across domains allows pre-allocation of vector capacity without per-task tuning.

\textit{LLM cost.} Using prompt templates in Appendix~\ref{appendix:chunking_prompt} and~\ref{appendix:query_prompt}, we estimate the token budget for a typical 200-token passage, and each passage will be segmented into 5 chunks, as follows:

\begin{itemize}
\item \textbf{Chunking + Coref}: 120-token system prompt + 200 input + $\sim$200 output $\Rightarrow$ 320 input / 200 output tokens (5 chunks);
\item \textbf{Pseudo-query Generation} (5 chunks): $5 \times$ (60 prompt + 40 input + 20 output) $\Rightarrow$ 500 input / 100 output tokens.
\end{itemize}

\noindent Totaling $\sim$820 input and 300 output tokens per passage ($\sim$1.2k tokens). With Mistral-Large-24.11 pricing of \$2/M input and \$6/M output tokens,\footnote{\href{https://mistral.ai/products/la-plateforme\#pricing}{https://mistral.ai/products/la-plateforme\#pricing}, last accessed 1 May 2025.}the estimated cost is:

\[
\begin{aligned}
\text{Cost} &= \$0.000002 \times (320+500) + \$0.000006 \times (200+100) \\
            &\approx \mathbf{\$0.0034} \text{ per passage}.
\end{aligned}
\]

\noindent Thus, preprocessing the 484k MS MARCO dev passages incurs a one-time cost of approximately \$1.6k. While this cost is non-trivial, it is incurred entirely offline and introduces no additional latency at retrieval time.

\section{Conclusion}

Effective retrieval often demands both global topical alighment and localized semantic precision, in this paper, We present \textbf{CLAP}, a lightweight retrieval augmentation framework that enhances first-stage retrieval by generating localized pseudo-queries from coreference-resolved semantic chunks. These queries effectively capture subtopic-level relevance signals, while the original passage representation preserves global alignment with the pre-trained dense retriever’s semantic space. A simple interpolation mechanism fuses these two views, balancing coarse and fine-grained relevance.

Experiments demonstrate that CLAP consistently improves retrieval effectiveness across both sparse and dense backbones. Notably, on out-of-domain datasets, CLAP boosts the upper bound of strong dense retrievers and, in some cases, matches or exceeds second-stage rerankers like BM25 + MonoT5-3B—highlighting its effectiveness in addressing domain shift. Moreover, CLAP maintains consistent performance gains regardless of the underlying retriever strength, demonstrating that its localized relevance signals complement rather than duplicate the semantics encoded by powerful dense models.

We attribute CLAP’s strong performance on out-of-domain datasets to its logic-intensive rather than knowledge-intensive design. Unlike expansion methods that rely on the language model’s pre-trained domain knowledge—which often limits generalization under distribution shift—CLAP performs a systematic transformation of each passage into logically coherent semantic chunks, explicitly resolves coreference relations, and generates localized pseudo-queries grounded in textual structure. This design enables robust generalization even when the language model lacks specialized domain knowledge, thereby improving retrieval effectiveness across diverse scenarios.

\section*{Limitations}

While our method demonstrates strong effectiveness and generalization across diverse retrieval settings, it has several notable limitations.

CLAP requires an offline pipeline that performs semantic chunking, coreference resolution, and pseudo-query generation before retrieval. It incurs substantial LLM-related preprocessing cost, especially when applied at large scale corpus. As analyzed in Section~\ref{analysis}, processing each 200-token passage consumes about 1.2k tokens of LLM usage, resulting in a non-trivial cost per passage—even though it is a one-time expense. Furthermore, the resulting index expands by $20$–$32\times$ due to the large number of generated pseudo-queries, leading to increased vector storage and retrieval computation overhead. Although retrieval remains linear in complexity, this footprint may pose challenges for deployment in low-resource or large-scale settings.

Moreover, CLAP's current fusion strategy uses a dataset-level static fusion weight $\alpha$ to balance global and local relevance signals. However, our experiments reveal that the optimal $\alpha$ varies across datasets and query types. This suggests that a more effective approach would involve query-adaptive fusion, where $\alpha$ is dynamically determined based on the characteristics of each query. 

Lastly, our method follows a multi-stage sequential pipeline where errors in earlier stages (e.g., inaccurate chunking or incomplete coreference resolution) can propagate downstream and degrade the quality of generated pseudo-queries. Although the use of a strong LLM mitigates some of these risks, fully robustifying each stage—especially under imperfect LLM outputs—remains an open challenge.

\section*{Acknowledgments}
\textbf{Use of Generative AI.} Portions of the manuscript (light copy-editing and LaTeX formatting suggestions) were assisted by ChatGPT (OpenAI; model: GPT\mbox{-}o1; accessed Feb--May 2025). The authors reviewed and verified all AI-assisted text and are solely responsible for the content.

\section{Appendices}
\appendix

\section{Prompt Template for Semantic Chunking and Coreference Resolution}
\label{appendix:chunking_prompt}

\begin{lstlisting}[basicstyle=\ttfamily\tiny,breaklines=true]
You are an expert linguistic analyst specialized in semantic decomposition, content segmentation, and coreference resolution.

Given the following text passage, your task is to identify and clearly separate its main semantic components into distinct chunks. **Importantly**, when creating each chunk, you must explicitly resolve pronouns or references (such as "it", "they", "this method", etc.) to their original and explicit noun entities to ensure semantic clarity and independence of each chunk.

### Instructions:

1. Carefully read and understand the passage.
2. Identify the key semantic subtopics or main conceptual components.
3. Divide the passage into multiple chunks, each chunk capturing exactly one coherent subtopic or conceptual unit. Each chunk must be semantically homogeneous and concise.
4. **Coreference Resolution**: In each chunk, explicitly replace pronouns or vague references with their original noun entities from the passage. Each chunk should stand alone with no ambiguity or unclear references.
5. For each chunk, provide a short descriptive title summarizing its main semantic content.
6. Return your response as a structured JSON array, with each object having three fields:
   - "chunk_id": incremental identifier (e.g., "a", "b", ...)
   - "chunk_title": a concise descriptive title of the chunk.
   - "chunk_text": the exact text of the chunk excerpted from the original passage, **with pronouns explicitly replaced by their original noun entities**.
7. If you cannot extract meaningful chunks, deem the passage as single chunk.

### Output format:
```json
[
  {
    "chunk_id": "a",
    "chunk_title": "title of first chunk",
    "chunk_text": "text of first chunk, pronouns explicitly replaced by original entities"
  },
  {
    "chunk_id": "b",
    "chunk_title": "title of second chunk",
    "chunk_text": "text of second chunk, pronouns explicitly replaced by original entities"
  }
  // Additional chunks if applicable...
]
Below are some examples (passage + structured output):

passage: Honey is primarily composed of sugars, mostly fructose and glucose. Bees produce honey by collecting nectar from flowers and storing it in honeycombs. It serves as a food source for bees during winter and has antibacterial properties beneficial for human health.

Structured Output: [ {"chunk_id": "a", "chunk_title": "Composition of Honey", "chunk_text": "Honey is primarily composed of sugars, mostly fructose and glucose."}, {"chunk_id": "b", "chunk_title": "How Bees Produce Honey", "chunk_text": "Bees produce honey by collecting nectar from flowers and storing it in honeycombs."}, {"chunk_id": "c", "chunk_title": "Functions and Benefits of Honey", "chunk_text": "Honey serves as a food source for bees during winter and has antibacterial properties beneficial for human health."} ]

passage: Photosynthesis is a process used by plants and other organisms to convert light energy into chemical energy.

Structured Output: [ {"chunk_id": "a", "chunk_title": "Definition of Photosynthesis", "chunk_text": "Photosynthesis is a process used by plants and other organisms to convert light energy into chemical energy."} ]

passage: Coffee beans grow primarily in tropical climates. They require specific conditions including adequate rainfall, temperatures between 15C and 24C, and shaded environments. The largest coffee-producing countries include Brazil, Vietnam, Colombia, and Indonesia. Coffee cultivation has significant economic importance, employing millions globally. However, coffee production can negatively impact the environment due to deforestation and high water usage.

Structured Output: [ {"chunk_id": "a", "chunk_title": "Optimal Climate for Coffee Cultivation", "chunk_text": "Coffee beans grow primarily in tropical climates. Coffee beans require specific conditions including adequate rainfall, temperatures between 15C and 24C, and shaded environments."}, {"chunk_id": "b", "chunk_title": "Major Coffee-Producing Countries", "chunk_text": "The largest coffee-producing countries include Brazil, Vietnam, Colombia, and Indonesia."}, {"chunk_id": "c", "chunk_title": "Economic Importance of Coffee Cultivation", "chunk_text": "Coffee cultivation has significant economic importance, employing millions globally."}, {"chunk_id": "d", "chunk_title": "Environmental Impacts of Coffee Production", "chunk_text": "However, coffee production can negatively impact the environment due to deforestation and high water usage."} ]

### Task Begin

passage:
Your passage text goes here.

\end{lstlisting}

\section{Prompt Template for Pseudo-query Generation}
\label{appendix:query_prompt}

\begin{lstlisting}[basicstyle=\ttfamily\tiny,breaklines=true]
You are an expert query generation system. Your task is to thoroughly analyze a given chunk of text along with its associated title. Based on these inputs, generate diverse user-like questions (pseudo-queries) from multiple distinct semantic aspects.

### Instructions:
1. Carefully read the provided inputs:
    - Title: The title or main subject of the document or passage.
    - Chunk: A short excerpt from the passage.

2. Generate additional distinct pseudo-queries:
    - Each pseudo-query must address a unique semantic aspect or perspective that can be explicitly answered by the given chunk and its title.

3. Structured fields for each pseudo-query:
    - "pseudo_query": The exact generated user-like question.

### Output format - Return the json only:
```json
[
  { "pseudo_query": "Do I need to pay taxes for a business with no income?" },
  // Additional pseudo-queries if applicable...
]
Below is an example (title + chunk + structured output):

title: IRS Tax Obligations on Nonexistent Income

chunk: I think you are going to find out that there are no taxes owed to the IRS for this nonexistent activity.

Structured Output: [ { "pseudo_query": "Do I need to pay taxes for a business with no income?" }, { "pseudo_query": "What are the IRS requirements for reporting nonexistent income?" }, { "pseudo_query": "Why don't I owe taxes for a business that didn't make any money?" }, { "pseudo_query": "How does the IRS handle tax filings for businesses with zero income?" } ] 

### Task Begin

title: {{ title }}
chunk: {{ chunk }}

\end{lstlisting}

\section{Descriptive Statistics of Similarity Gain}


\label{appendix:similarity_stats}
\begin{table}[H]
\centering
\small
\setlength{\tabcolsep}{4pt}
\renewcommand{\arraystretch}{1.2}
\resizebox{\linewidth}{!}{
\begin{tabular}{l|cccccccccccc}
\toprule
\textbf{Dataset} & Mean & Std Dev & Var & Min & 10\%Q & 25\%Q & Median & 75\%Q & Max & Skew & Kurt & Std/Mean \\
\midrule
FiQA      & 0.2165 & 0.1667 & 0.0278 & -0.1771 & 0.0278 & 0.0976 & 0.1905 & 0.3157 & 0.9394 & 0.7278 & 0.7518 & 0.7698 \\
NFCorpus  & 0.1232 & 0.0776 & 0.0060 & -0.2711 & 0.0318 & 0.0737 & 0.1205 & 0.1689 & 0.5468 & 0.2921 & 1.3929 & 0.6299 \\
\bottomrule
\end{tabular}
}
\caption{Descriptive statistics of similarity gain for FiQA and NFCorpus on all-MiniLM-L6-v2.}
\end{table}

\section{nDCG@10 vs $\alpha$ across datasets}
\label{appendix:appendix_alpha}
\vspace{-1em}
\begin{figure}[H]
    \centering
    \includegraphics[width=\linewidth]{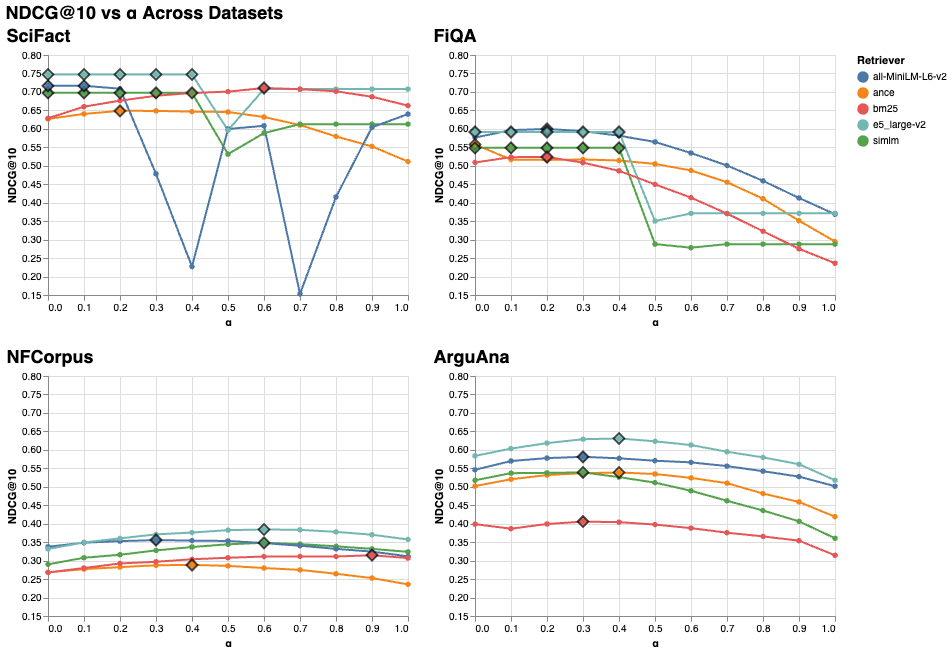}
    \caption{\textbf{Effect of fusion weight $\alpha$ on retrieval performance (NDCG@10)} across datasets and retrievers.
    Larger $\alpha$ favors global query–passage matching, while smaller $\alpha$ emphasizes local pseudo-query signals.
    Diamond markers indicate the optimal $\alpha$ selected per retriever per dataset.}
    \label{fig:viz-alpha}
\end{figure}

\newpage
\bibliographystyle{ACM-Reference-Format}
\bibliography{clap}

\end{document}